\documentclass[prd,aps,10pt,nofootinbib,twocolumn,superscriptaddress,preprintnumbers,balancelastpage,longbibliography]{revtex4-1}

\usepackage{amsmath,amssymb}	
\usepackage{mathtools}
\usepackage{fontawesome}
\usepackage[dvipsnames]{xcolor}
\usepackage{hyperref}
\usepackage{xspace}
\usepackage{fancyhdr}
\usepackage{graphicx}
\usepackage{siunitx}

\definecolor{linkcolor}{rgb}{0.7752941176470588, 0.22078431372549023, 0.2262745098039215}

\newcommand{\nbicon}{{\color{linkcolor}\faFileCodeO}\xspace}
\newcommand{\nblink}[1]{\href{https://github.com/smsharma/edges-endpoints-21cm/notebooks/#1.ipynb}{\nbicon}}
\newcommand{\githubmaster}{\href{https://github.com/smsharma/edges-endpoints-21cm/}{\faGithub}\xspace}

\newcommand{\dd}{\mathrm{d}}
\newcommand{\mAp}{m_{A^\prime}}

\definecolor{deepgreen}{rgb}{0.2,0.8,0.2}

\hypersetup{colorlinks=true,
linkcolor=linkcolor,
citecolor=linkcolor,
urlcolor=linkcolor,
linktocpage=true,
pdfproducer=medialab,
}

\DeclareSIUnit \h {\ensuremath{\mathit{h}}}
\DeclareSIUnit\electronvolt{e\kern-.05em V}
\DeclareSIUnit\parsec{pc}

\begin{document}

\preprint{DESY 20-143}

\title{Edges and Endpoints in 21-cm Observations from Resonant Photon Production}

\author{Andrea Caputo}
\email{andrea.caputo@uv.es}
\thanks{ORCID: \href{https://orcid.org/0000-0003-1122-6606}{0000-0003-1122-6606}}
\affiliation{Instituto de F\'{i}sica Corpuscular, CSIC-Universitat de Valencia, Apartado de Correos 22085, E-46071, Spain}

\author{Hongwan Liu}
\email{hongwanl@princeton.edu}
\thanks{ORCID: \href{https://orcid.org/0000-0003-2486-0681}{0000-0003-2486-0681}}
\affiliation{Center for Cosmology and Particle Physics, Department of Physics, New York University, New York, NY 10003, USA}
\affiliation{Department of Physics, Princeton University, Princeton, NJ 08544, USA}

\author{Siddharth Mishra-Sharma}
\email{sm8383@nyu.edu}
\thanks{ORCID: \href{https://orcid.org/0000-0001-9088-7845}{0000-0001-9088-7845}}
\affiliation{Center for Cosmology and Particle Physics, Department of Physics, New York University, New York, NY 10003, USA}

\author{Maxim Pospelov}
\affiliation{School of Physics and Astronomy, University of Minnesota, Minneapolis, MN 55455, USA}
\affiliation{William I. Fine Theoretical Physics Institute, School of Physics and Astronomy, University of Minnesota, Minneapolis, MN 55455, USA}

\author{Joshua T. Ruderman}
\email{ruderman@nyu.edu}
\thanks{ORCID: \href{https://orcid.org/0000-0001-6051-9216}{0000-0001-6051-9216}}
\affiliation{Center for Cosmology and Particle Physics, Department of Physics, New York University, New York, NY 10003, USA}
\affiliation{Deutsches Elektronen-Synchrotron (DESY), D-22607 Hamburg, Germany}

\author{Alfredo Urbano}
\email{alfredo.urbano@sissa.it}
\thanks{ORCID: \href{https://orcid.org/0000-0002-0488-3256}{0000-0002-0488-3256}}
\affiliation{INFN sezione di Trieste, SISSA, via Bonomea 265, I-34132 Trieste, Italy}
\affiliation{IFPU, Institute for Fundamental Physics of the Universe, via Beirut 2, I-34014 Trieste, Italy}

\date{\protect\today}

\begin{abstract}
We introduce a novel class of signatures---spectral edges and endpoints---in 21-cm measurements resulting from interactions between the standard and dark sectors. Within the context of a kinetically mixed dark photon, we demonstrate how resonant dark photon-to-photon conversions can imprint distinctive spectral features in the observed 21-cm brightness temperature, with implications for current, upcoming, and proposed experiments targeting the cosmic dawn and the dark ages. These signatures open up a qualitatively new way to look for physics beyond the Standard Model using 21-cm observations. \githubmaster
\end{abstract}

\maketitle

\noindent
{\bf Introduction.---}Observation of the redshifted 21-cm line emission from neutral hydrogen in the intergalactic medium (IGM) has recently emerged as a powerful probe of the cosmological history of our Universe. The intensity of the global 21-cm emission can be measured as the differential brightness temperature of the hydrogen spin temperature contrasted against the background radiation and scales roughly as $\Delta T_\mathrm{b} \propto x_\mathrm{HI}(1 - T_\gamma/T_\mathrm{s})$, where $x_\mathrm{HI}$ is 
the  neutral  hydrogen fraction, $T_\mathrm{s}$ is the transition spin temperature, and $T_\gamma$ is the temperature of the background radiation. Its sensitive dependence on the underlying radiation fields as well as cosmic heating and ionization processes makes it a powerful probe of astrophysics as well as physics beyond the Standard Model (SM).

In the standard cosmological picture, the formation of the first stars and galaxies during cosmic dawn couples the spin and kinetic temperatures via the absorption and re-emission of Lyman-$\alpha$ photons (the Wouthuysen-Field effect~\cite{1952AJ.....57R..31W,1959ApJ...129..536F}), producing a distinctive absorption trough in the observed 21-cm brightness temperature as the spin temperature cools. Eventually, X-ray sources reheat the gas, and ultraviolet radiation emitted by stellar sources leads to reionization, increasing the kinetic temperature and turning off the absorption feature. At higher redshifts, radiative coupling of the spin and photon temperatures leads to $\Delta T_\mathrm{b} \sim 0$, although the decoupling of the photon and kinetic temperatures around $z\sim150$ and collisional coupling of the spin and kinetic temperatures leads to a minor absorption trough around $z\sim100$. While the depth of the absorption feature at cosmic dawn depends sensitively on the assumed astrophysics and cosmology, a bound on the maximal absorption is obtained by taking the limit of perfect coupling of the spin and kinetic temperatures and exclusively adiabatic cooling, which in the standard $\Lambda$CDM scenario corresponds to $\Delta T_\mathrm{b}(z=17) \approx -0.2\,\mathrm{K}$.

The EDGES experiment recently reported the measurement of the first global 21-cm signal at cosmic dawn~\cite{Bowman:2018yin}, with a central value $\Delta T_\mathrm{b}(z=17) \simeq -0.5^{+0.2}_{-0.5}\,\mathrm{K}$ at 99\% confidence level which, when taken at face value, implies a $\sim3.8\sigma$ disagreement with the minimum allowed value in the standard scenario. In addition to the depth, the shape of the absorption signal as measured by EDGES is unexpected as well, with the sharp turn-on and turn-off implying sudden Lyman-$\alpha$ injection and then sudden heating during reionization, contrary to expectation from more standard astrophysical scenarios. Taking these tensions at face value would imply the need for a modification to the standard cosmology. 

\begin{figure*}[!t]
\centering
\includegraphics[width=0.47\textwidth]{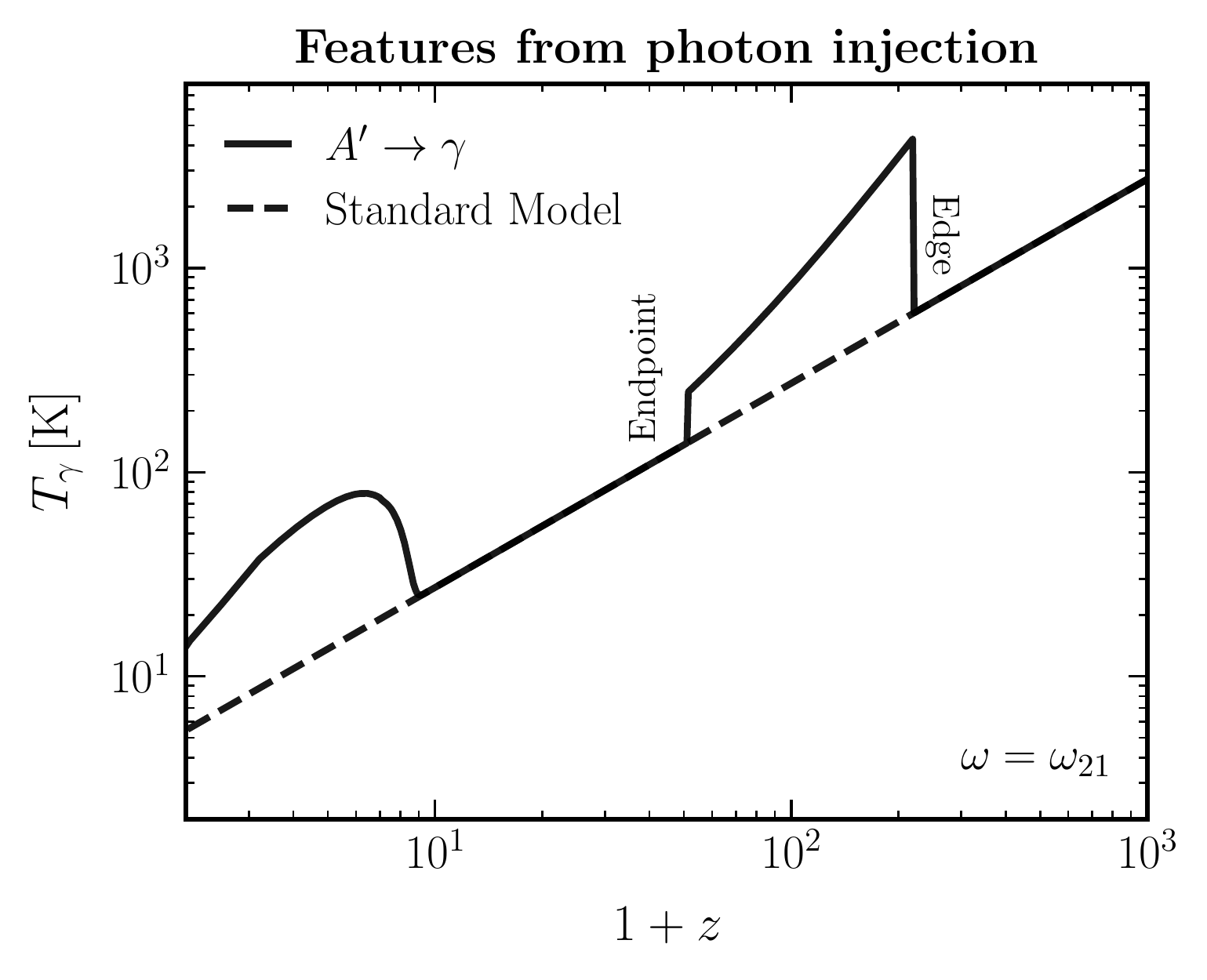}
\includegraphics[width=0.47\textwidth]{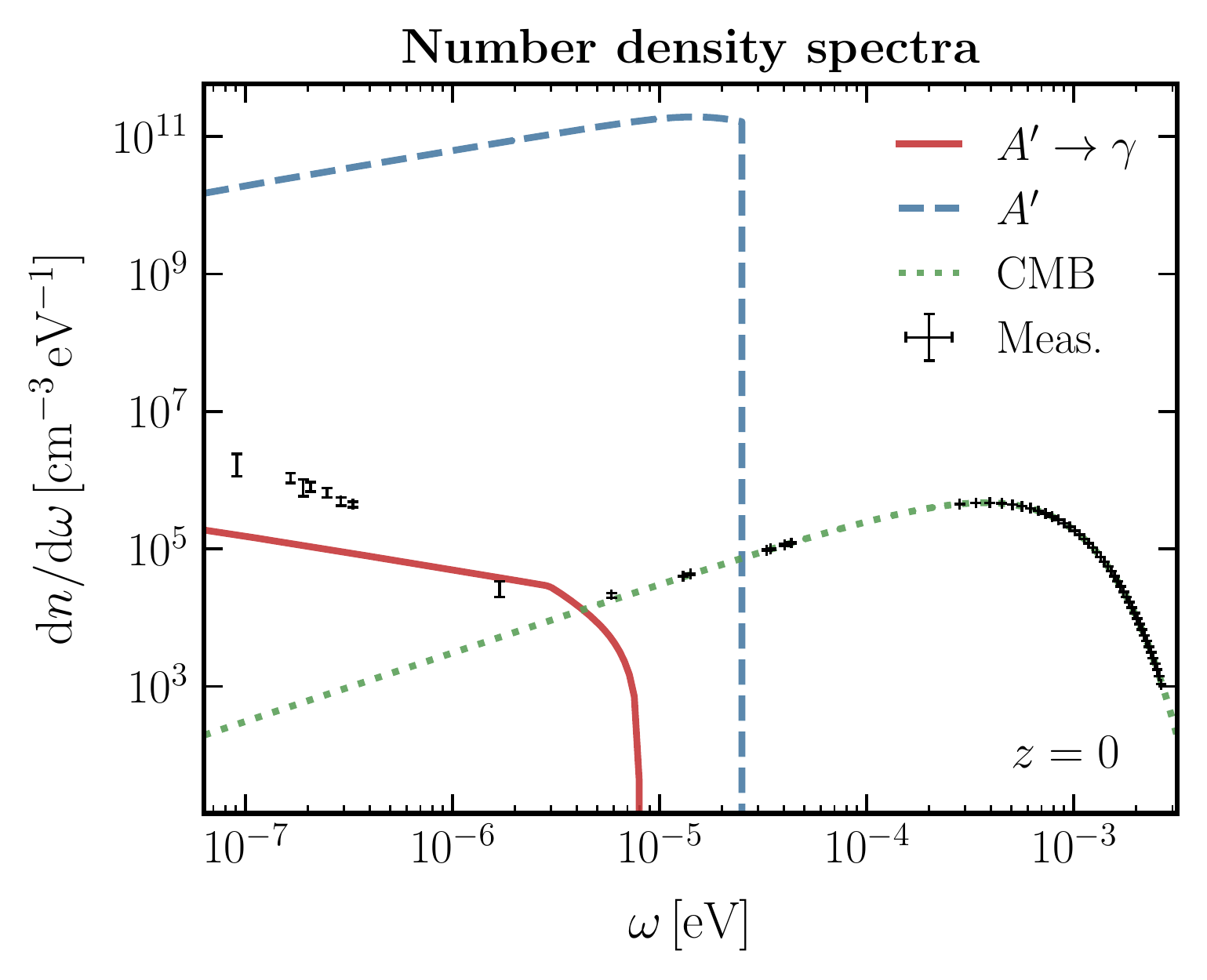}
\caption{(Left) Evolution of the 21-cm photon temperature for the standard case (dashed black) and including photon injection for an illustrative parameter point with dark photon mass $m_{A^{\prime}} = 10^{-12}\,\mathrm{eV}$, DM mass $m_a = 5.1\times10^{-5}\,\mathrm{eV}$, and kinetic mixing $\epsilon=1.4\times10^{-9}$. A sharp increase in effective temperature at $z\simeq 220$ followed by a turn-off at $z\simeq 50$ (corresponding to the regime where all injected photons have redshifted below 21-cm) can be seen, along with additional resonant injection at lower redshifts $z\lesssim 10$. (Right) For the same parameter point, the present-day differential number density spectra for dark photons (dashed blue), corresponding resonantly injected photons (solid red), and the standard CMB (dashed green) along with measurements from COBE/FIRAS~\cite{Fixsen:1996nj}, ARCADE2~\cite{Fixsen:2009xn}, and LWA radio surveys~\cite{Dowell:2018mdb} (black data points). A large excess in the photon number density in the RJ tail is consistent with observations of the CMB spectrum.}
\label{fig:features}
\end{figure*}

Several mechanisms, both astrophysical and those invoking physics beyond the SM, have been proposed to explain the EDGES observation, which would necessarily imply a larger differential in the photon and spin temperatures compared to the standard expectation~\cite{Feng:2018rje,Barkana:2018lgd,Berlin:2018sjs,Barkana:2018cct,Fraser:2018acy,Pospelov:2018kdh,Moroi:2018vci,Fialkov:2019vnb,Liu:2019knx,Choi:2019jwx}. Injection of Lyman-$\alpha$ photons from the most massive halos~\cite{Kaurov:2018kez} and efficient star-formation along with excess X-ray heating~\cite{Mirocha:2018cih} are examples of non-standard astrophysical explanations for the sharp turn-on and turn-off of the absorption feature. Mechanisms for cooling of the kinetic temperature beyond adiabatic cooling via relative velocity-dependent interaction between baryons and dark matter (DM) particles with kinetic temperature below the IGM temperature have been proposed~\cite{Barkana:2018lgd}, strongly constrained~\cite{Barkana:2018cct, Berlin:2018sjs}, and recently revived~\cite{Liu:2019knx}. Another class of explanations relies on raising the effective radio background temperature beyond the standard temperature of the Cosmic Microwave Background (CMB), $T_\gamma(z) = T_\mathrm{CMB, 0}(1+z)$, where $T_\mathrm{CMB, 0}\approx2.725\,\mathrm{K}$ is the present-day CMB temperature. In particular, Ref.~\cite{Pospelov:2018kdh} proposed raising the effective temperature through the production and subsequent resonant oscillation of dark photons into SM photons in the Rayleigh-Jeans (RJ) tail of the CMB\@. This scenario was further explored in Ref.~\cite{Moroi:2018vci} in the context of dark radiation consisting of axion-like particles (ALPs) and in Ref.~\cite{Choi:2019jwx} in the context of ALP-photon-dark photon oscillations in the presence of a primordial dark magnetic field. The basic idea is that the decay of cosmologically long-lived dark sector particles making up a large fraction of the DM density with masses in the meV range into dark photons can result in a much larger number density of dark photons in the RJ tail of the CMB as compared to regular photons. The subsequent resonant conversion of these dark photons into SM photons via a mechanism such as kinetic mixing~\cite{Holdom:1985ag} can enhance the number density of RJ photons and result in a deeper 21-cm absorption feature. 

In this Letter, we study the distinctive ways resonant photon injection can imprint itself onto a measured global 21-cm signal. In particular, we showcase scenarios in which spectral features imprinted through resonant photon production can naturally explain the depth and shape of the measured EDGES absorption feature and discuss implications for constraining these scenarios with future 21-cm measurements. We describe for the first time characteristic spectral features---edges and endpoints---in measurements of 21-cm photons sourced during the cosmic dark ages~\cite{Alvarez:2019pss, Burns:2019zia, Koopmans:2019wbn} and originating from coupling ordinary photons to particles of the dark sector. These novel signatures have the potential to be powerful probes of physics beyond the SM\@. 

Throughout this work, we use units with $\hbar = c = k_\mathrm{B} = 1$, and the \textit{Planck} 2018 cosmology~\cite{Aghanim:2018eyx}. For reproducibility, code used to produce the results in this Letter is available on GitHub~\githubmaster.

\noindent
{\bf Spectral features due to photon injection.---}Although photon injection can arise in a variety of models~\cite{Pospelov:2018kdh, Moroi:2018vci, Choi:2019jwx}, for concreteness we focus on the scenario introduced in Ref.~\cite{Pospelov:2018kdh} where a cosmologically long-lived dark sector particle $a$ of mass $m_a$ with lifetime $\tau_a$ decays into dark photons $A'$ of mass $m_{A^\prime}$ through $a \rightarrow A'A'$, which subsequently resonantly convert into regular photons, $A'\to \gamma$, when their mass matches the photon plasma mass $m_\gamma(\vec x,z)$~\cite{Mirizzi:2009iz,Caputo:2020bdy,Caputo:2020rnx}. The conversion results in a sharp increase in the number density of photons in the RJ tail of the CMB, which contribute to the 21-cm background photon temperature. The redshift of this feature, which we call an ``edge'', is around
\begin{equation} \label{eq:edge}
z_\mathrm{edge} =z_\mathrm{res};~~~~~\overline{m_\gamma(z_\mathrm{res})} = \mAp,
\end{equation}
where $z_\mathrm{res}$ is the resonance redshift at which the plasma and dark photon masses match and $\overline{m_\gamma(z_\mathrm{res})}$ is the mean plasma mass at that redshift. This results in a near-instantaneous increase in the effective photon temperature, (further) decoupling the spin and photon temperatures, a consequence of which is an enhancement of the 21-cm brightness absorption feature.  Measuring the location of the edge uniquely determines the dark photon mass, through Eq.~\eqref{eq:edge}.

Photons resonantly produced at a given redshift $z_\mathrm{res}$ and frequency $\omega_\mathrm{res}$ then evolve to contribute to the number density of 21-cm photons at a given redshift $z_{21}$ as $\omega_\mathrm{res}(1 + z_{21}) = \omega_{21}(1 + z_\mathrm{res})$. Kinematically, $\omega_\mathrm{res} > m_a/2$ is forbidden for a two-body decay, resulting in a spectral feature, which we call an ``endpoint'', beyond which all of the converted photons have redshifted below the 21-cm frequency. The location of the endpoint $z_{\mathrm{end}}$ is defined through
\begin{equation} \label{eq:endpoint}
\frac{1+z_{\mathrm{end}}}{1+z_{\mathrm{res}}}=\frac{\omega_{21}}{m_{a} / 2}.
\end{equation}
Measuring both the location of the edge and endpoint uniquely determines both $m_{A'}$ and $m_a$, through Eqs.~\eqref{eq:edge} and~\eqref{eq:endpoint}.

We note that the edges and endpoints that we identify in 21-cm are analogues to the edges and endpoints that can signify new physics in kinematic distributions at high energy colliders~\cite{Hinchliffe:1996iu}.  In both cases, these distinctive spectral features serve as handles to distinguish new physics from backgrounds.

For the remainder of this Letter we assume that $a$ is the DM\@.  The differential number density of dark photons of angular frequency $\omega$ at redshift $z$ due to $a$ decays is given by~\cite{Pospelov:2018kdh,Cui:2017ytb,Garcia:2020qrp}
\begin{equation}
\frac{\dd n_{A'}}{\dd \omega}=\frac{2 \rho_{\mathrm{DM}}\left(z_{\mathrm{dec}}\right)(1 + z)^3}{\tau_a H\left(z_{\mathrm{dec}}\right) m_{a} \omega\left(1+z_{\mathrm{dec}}\right)^{3}} \Theta\left(\frac{m_{a}}{2}-\omega\right),
\label{eq:dn_domega}
\end{equation}
where $z_{\mathrm{dec}}$ is the redshift at which the decay $a \rightarrow A'A'$ takes place, 
$H\left(z_{\mathrm{dec}}\right)$ the Hubble rate and 
 $\rho_{\mathrm{DM}}\left(z_{\mathrm{dec}}\right)$ the DM density, both evaluated at $z_{\mathrm{dec}}$.
Eq.~\eqref{eq:dn_domega} presumes 2-body decay kinematics with $\tau_a \gg t_\mathrm{U}$, and takes $m_a\gg m_{A'}$. 
Note that for a 2-body decay there exists a unique relation $z_{\mathrm{dec}} =m_a/(2\omega) - 1$ between angular frequency and redshift of decay.
The  photon abundance $\dd n_{\gamma}/\dd \omega$ is obtained by multiplying $\dd n_{A'}/\dd \omega$ by the total $A'\rightarrow \gamma$ conversion probability $\left\langle P_{\gamma \rightarrow A^{\prime}}\right\rangle = \int_z^{z_\mathrm{dec}}\dd z'\,{\dd \left\langle P_{\gamma \rightarrow A^{\prime}}\right\rangle}/{\dd z'}$, computed following Refs.~\cite{Caputo:2020bdy,Caputo:2020rnx} and accounting for effects of inhomogeneities in the plasma mass. Note that since perturbations in the plasma mass result in resonant conversion over some range of redshifts around $z_\mathrm{edge}$~\cite{Caputo:2020bdy,Caputo:2020rnx,Bondarenko:2020moh,Garcia:2020qrp,Witte:2020rvb}, it will also take different amounts of time for their wavelengths to redshift beyond that corresponding to 21-cm. This results, in general, in the spectral endpoint having a characteristic width. We use the fiducial setup from Ref.~\cite{Caputo:2020bdy}, with a log-normal description of plasma mass fluctuations and a simulation-inferred fluctuation spectrum~\cite{Caputo:2020rnx} to compute conversion probabilities accounting for plasma mass perturbations. As in Ref.~\cite{Caputo:2020bdy}, we only consider plasma mass perturbations in the range $10^{-2} < 1 + \delta < 10^2$ throughout this work, since the log-normal distribution of perturbations cannot be assumed to be reliable outside of this range~\cite{Caputo:2020bdy,Caputo:2020rnx}.

The effective 21-cm photon temperature including the excess photons $n_{A'\rightarrow\gamma}$ is computed as $T_\gamma(z) = T_{\mathrm{CMB}}(z) (1 + n_{A'\rightarrow\gamma}/n_\mathrm{CMB})$, where $T_{\mathrm{CMB}}(z)$ is the standard CMB temperature and $n_\mathrm{CMB}$ the standard CMB number density corresponding to 21-cm.  This effective 21-cm temperature is shown in the left panel of Fig.~\ref{fig:features} for an illustrative signal point with dark photon mass $m_{A^{\prime}} = 10^{-12}\,\mathrm{eV}$, DM mass $m_a = 5.1\times10^{-5}\,\mathrm{eV}$, and kinetic mixing $\epsilon=10^{-9}$. This corresponds to resonant conversion around $z_\mathrm{res}\simeq 220$ and a kinematic endpoint around $z_\mathrm{end} \simeq 50$. An additional resonance at late times $z \lesssim 10$ is seen in this case due to conversions in overdense plasma regions during reionization~\cite{Caputo:2020bdy,Caputo:2020rnx}. 

Several constraints on this parameter space apply---\emph{(i)}~constraints from stellar energy loss due to $A'a$ pair production~\cite{Haft:1993jt,Pospelov:2018kdh}, \emph{(ii)}~constraints on excess $A'\rightarrow\gamma$ photon flux from radio and microwave observations~\cite{Fixsen:2009xn, Dowell:2018mdb}, \emph{(iii)}~constraints on $\gamma\leftrightarrow A'$ from COBE/FIRAS~\cite{Fixsen:1996nj,Caputo:2020bdy}, and~\emph{(iv)} bounds on the DM lifetime $\tau_a$. The DM lifetime throughout this study is chosen to saturate the allowed bound~\cite{Poulin:2016nat}. The right panel of Fig.~\ref{fig:features} shows the present-day number density spectrum of dark photons (dashed blue) and photons (solid red) compared to the standard CMB expectation (dotted green), along the measured values from COBE/FIRAS~\cite{Fixsen:1996nj}, ARCADE2~\cite{Fixsen:2009xn}, and LWA radio surveys~\cite{Dowell:2018mdb} for the illustrative parameter point. It can be seen that the photon spectrum in this case runs up against the measured radio flux at $\omega\simeq 2\times 10^{-6}$\,eV, constraining the maximum allowed injection.
Such an enhancement in the 21-cm photon temperature is allowed by current constraints and can lead to striking signatures observable by current and future 21-cm and radio surveys, opening up a new avenue for probing the dark sector. 

\begin{figure*}[!t]
\centering
\includegraphics[width=0.47\textwidth]{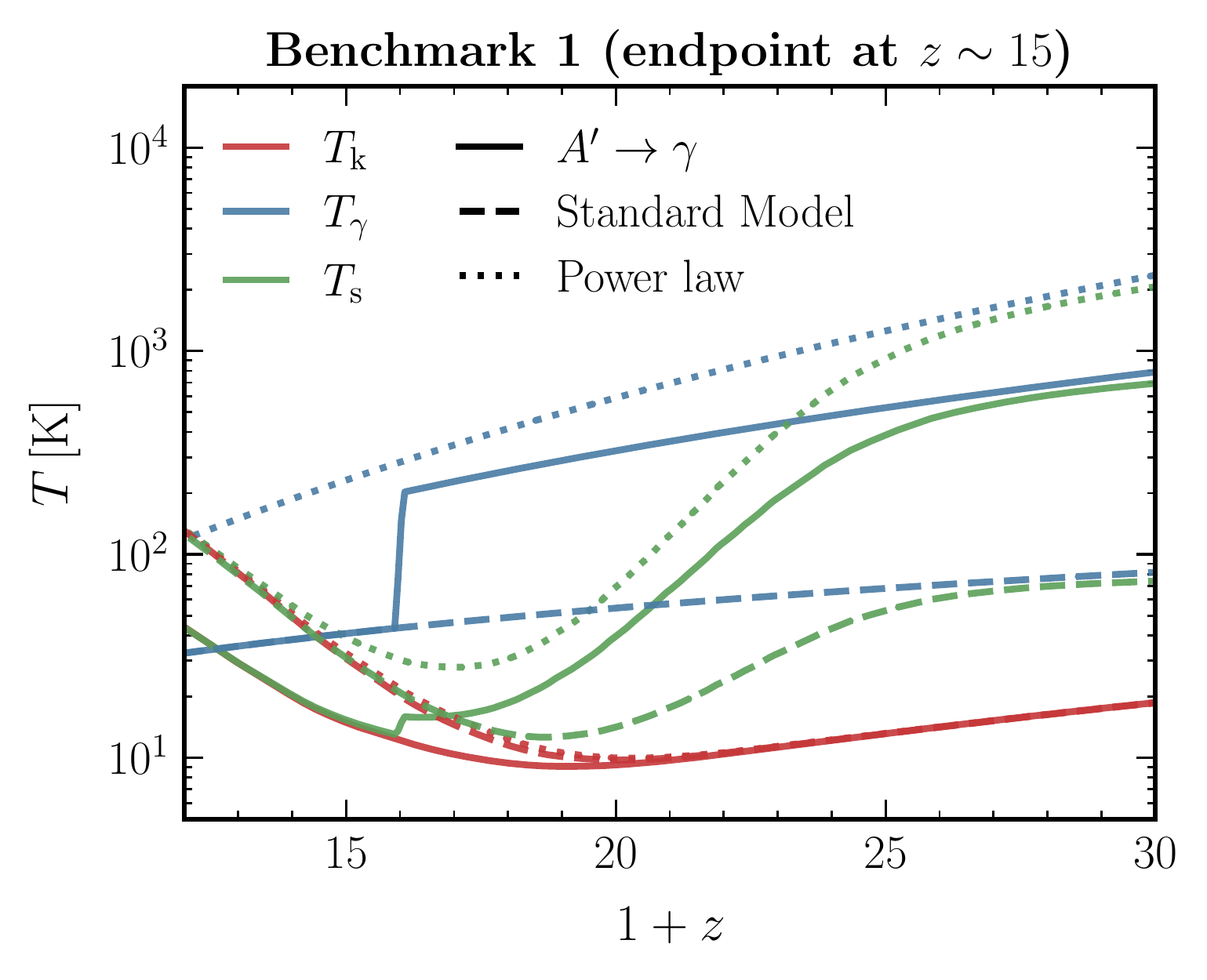}
\includegraphics[width=0.47\textwidth]{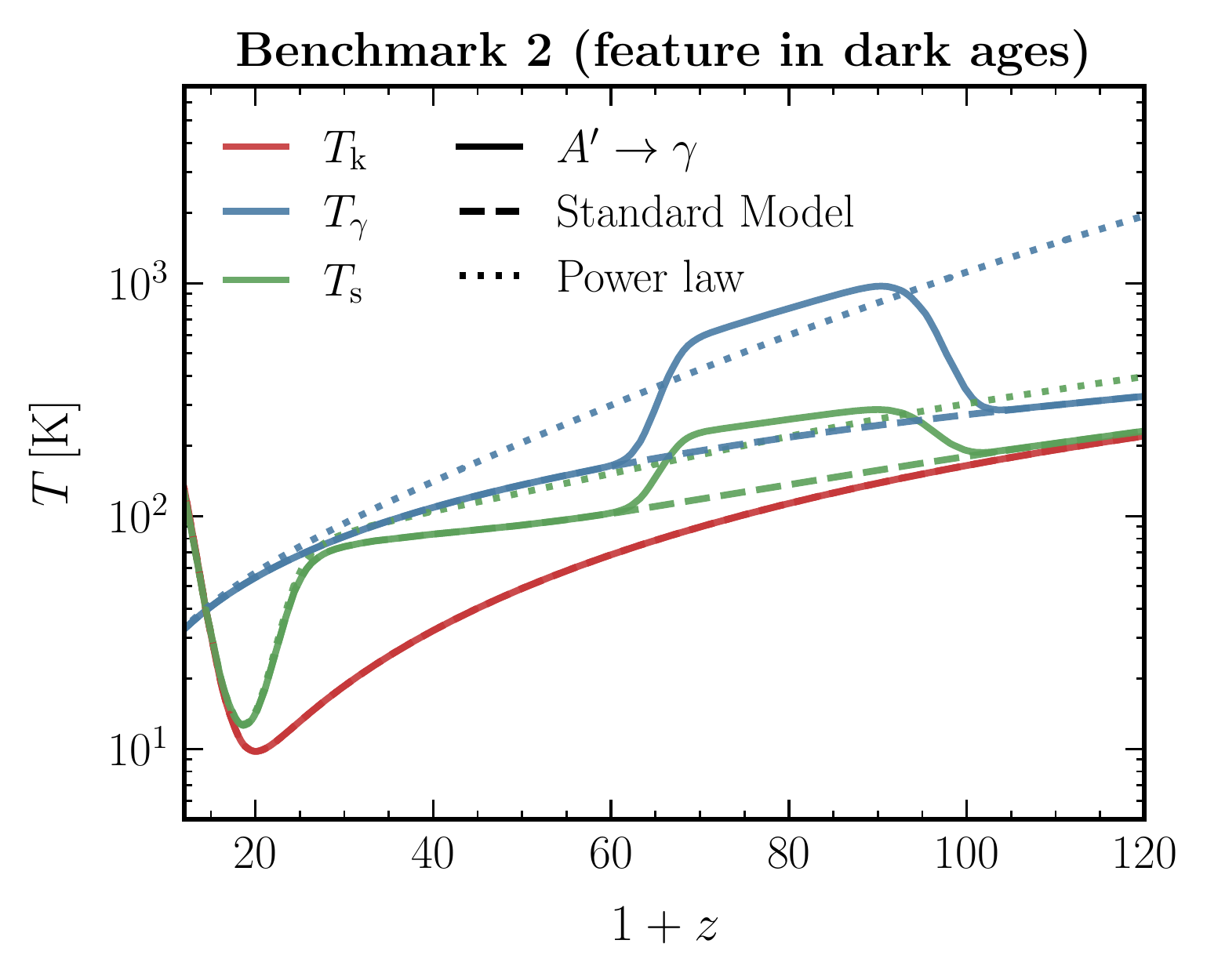}
\caption{Evolution of the kinetic (red), photon (blue), and spin (green) temperatures with redshift, shown for the Standard Model (dashed), phenomenological power law photon injection (dotted), and photon injection due to dark photon resonant conversion $A^\prime\rightarrow\gamma$ (solid) shown for Benchmark 1 (left) and Benchmark 2 (right). Compared to the Standard Model and power law cases, reduced X-ray heating is assumed for the $A^\prime\rightarrow\gamma$ scenarios.}
\label{fig:temperatures}
\end{figure*}

\noindent
{\bf Applications to 21-cm observations.---}The excess photon flux and, consequently, higher effective temperature resulting from resonant photon injection can leave characteristic imprints on the observed 21-cm signal. We focus on two benchmark scenarios here in order to illustrate the main qualitative features and relevance to current and future 21-cm measurements.

\noindent\emph{Benchmark 1: Spectral endpoint at $z\sim15$}---A sharp turn-off in the photon temperature evolution due to a spectral endpoint will decrease the contrast between the spin and photon temperatures, turning off the 21-cm absorption feature. This parameter point corresponds to dark photon mass $m_{A^{\prime}} = 10^{-11}\,\mathrm{eV}$, DM mass $m_a = 4.9\times10^{-4}\,\mathrm{eV}$, and kinetic mixing $\epsilon=5\times10^{-8}$, which would result in resonant conversion around $z_\mathrm{edge}\simeq 660$ and an endpoint around $z_\mathrm{end}\simeq 15$.

\noindent\emph{Benchmark 2: Spectral features during the dark ages $z\sim 50\text{--}95$}---An edge or endpoint during the dark ages would result in a spectral feature potentially detectable with proposed space-based 21-cm measurements~\cite{Alvarez:2019pss, Burns:2019zia, Koopmans:2019wbn}. This parameter point corresponds to dark photon mass $m_{A^{\prime}} = 2.5\times10^{-13}\,\mathrm{eV}$, DM mass $m_a = 1.7\times10^{-5}\,\mathrm{eV}$, and kinetic mixing $\epsilon=4.5\times 10^{-10}$, which would result in an edge around $z_\mathrm{edge}\simeq 95$ and a kinematic endpoint around $z_\mathrm{end} = 65$.

The evolution of the kinetic, photon, and spin temperatures for Benchmarks 1 and 2 is shown in the left and right panels of Fig.~\ref{fig:temperatures}, respectively. We employ the toy model for Lyman-$\alpha$ and X-ray heating~\cite{Hirata:2005mz} with additional input from Refs.~\cite{Venumadhav:2018uwn,Furlanetto:2009uf,Furlanetto:2009uf,Ali-Haimoud:2013hpa,Mesinger:2010ne} to compute the temperature evolution; details of our global 21-cm computation are described in App.~\ref{app:21-cm-code}\@. A halo virial temperature cut $T_\mathrm{vir} = 2\times10^4$\,K and star-formation efficiency $f_* = 3\%$ is assumed by default, with the effective X-ray star-formation efficiency for Benchmark 1 lowered to 1\% to demonstrate the effect of the spectral endpoint. For comparison, we also show the temperature evolution for the standard cosmological scenario with $T_\gamma=T_{\mathrm{CMB}, 0}(1+z)$ (dashed, labeled ``Standard Model'') and a photon injection with parameterized power law temperature evolution $T_\gamma=T_{\mathrm{CMB}, 0}(1+z)\left[1+f_\mathrm{r}A_{\mathrm{r}}\left(\nu_{0}/78\,\mathrm{MHz}\right)^{\beta}\right]$ (dotted, labeled ``Power law''), where $\nu_0$ is the present-day photon frequency, and $A_{\mathrm{r}}$ and $\beta$ are motivated by and fit to the excess low-frequency radio background measured by ARCADE2~\cite{Fixsen:2009xn} and LWA~\cite{Dowell:2018mdb} for $f_\mathrm{r} = 1$ as in Refs.~\cite{Fialkov:2019vnb, Feng:2018rje}. When comparing to Benchmark 1, we set $f_\mathrm{r} = 2\%$ in order to obtain an absorption depth consistent with the fiducial EDGES measurement. When comparing to Benchmark 2 on the other hand,  $f_\mathrm{r} = 0.01\%$ of the radio emission is chosen to illustrate its signature during the dark ages and compare with the resonant photon injection scenario. 

\begin{figure*}[!htbp]
\centering
\includegraphics[width=0.47\textwidth]{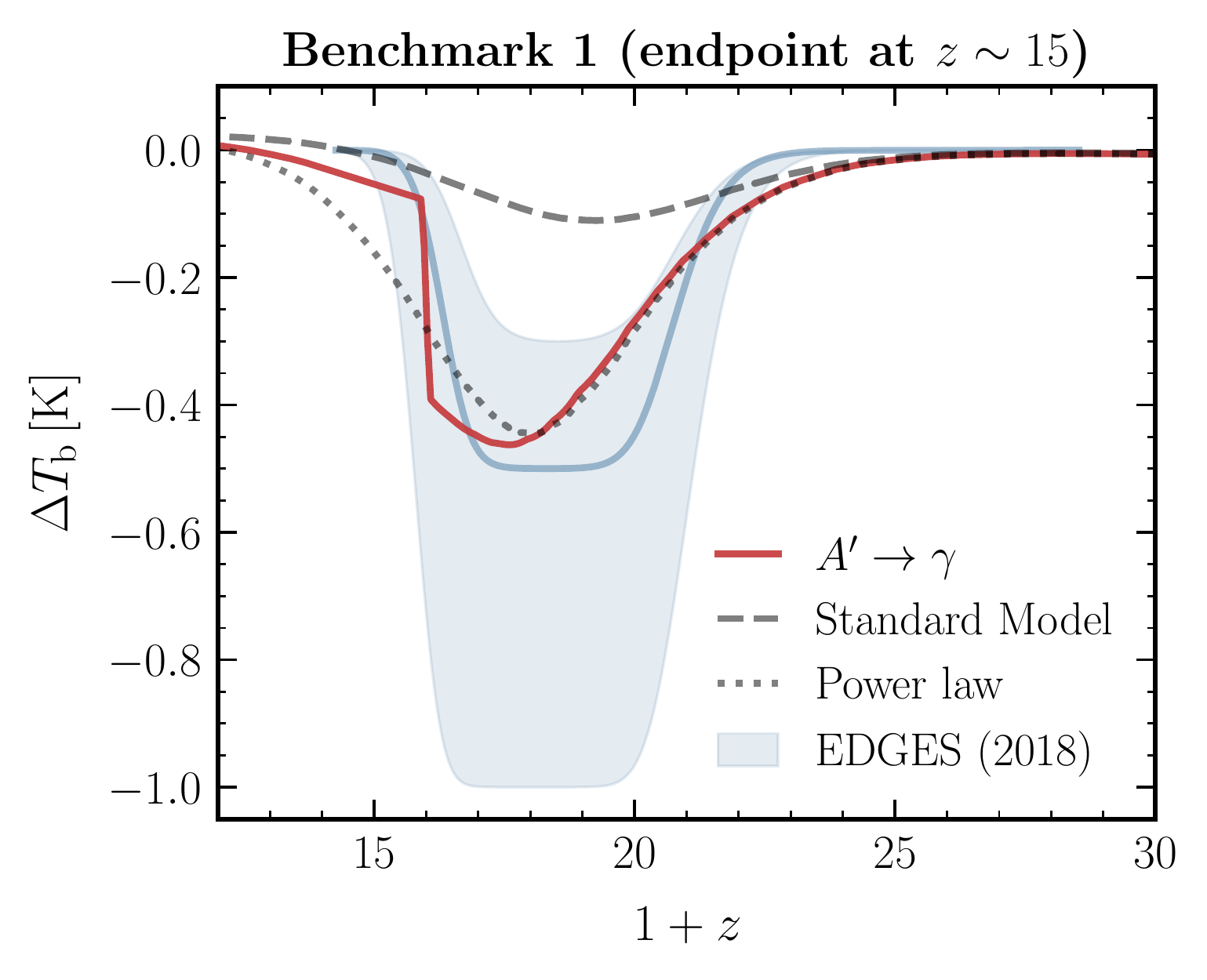}
\includegraphics[width=0.47\textwidth]{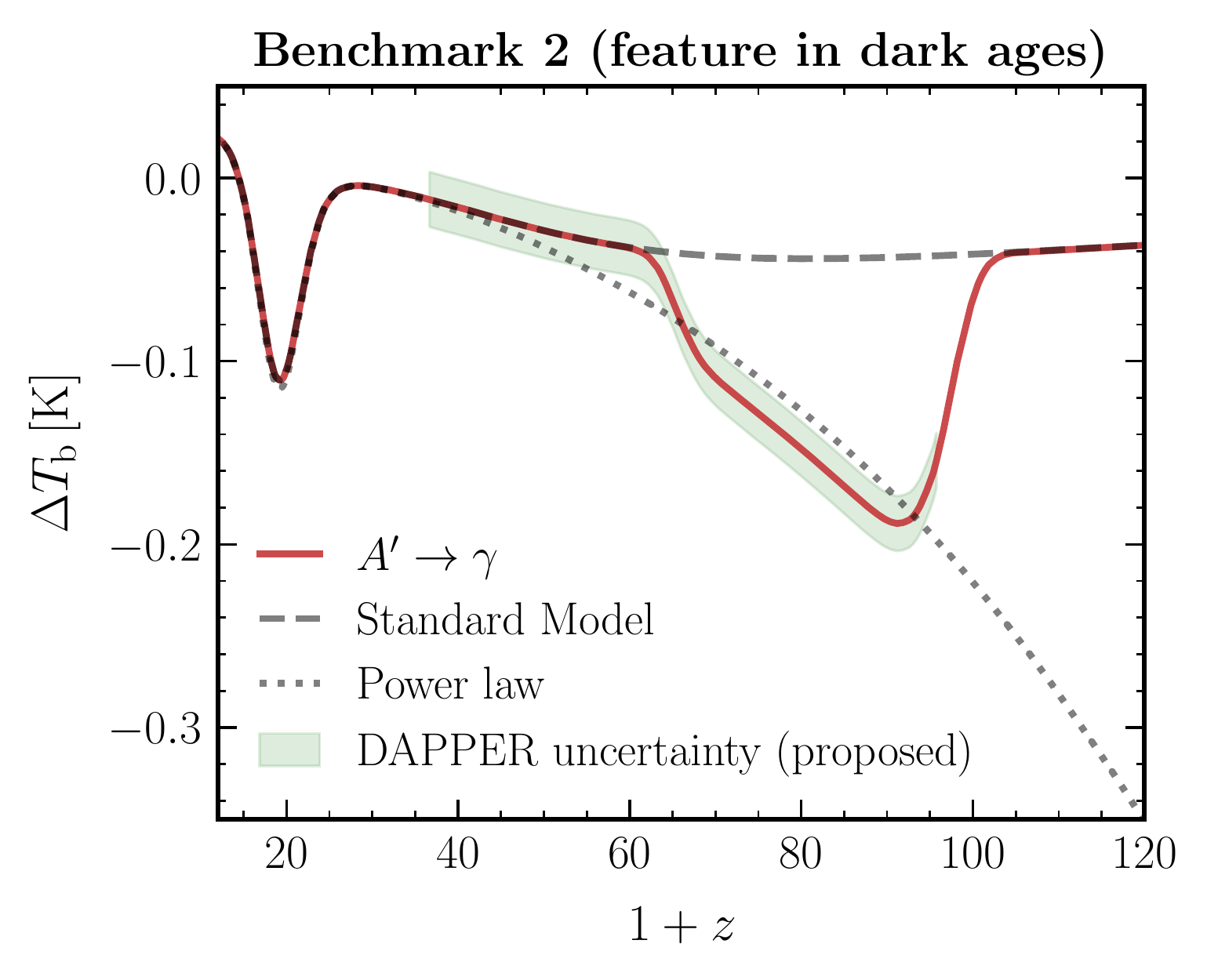}
\caption{The 21-cm brightness temperature contrast relative to the microwave background (solid red) for the benchmark scenarios considered: (left) Benchmark 1, showing a kinematic endpoint at $z=15$ and resulting in a sharp turnoff in the absorption feature that could explain the EDGES measurement (blue band), and (right) Benchmark 2, showing a distinctive kinematic feature during the dark ages. The expected 15\,mK uncertainty of the space-based DAPPER experiment in the 15--40\,MHz range~\cite{Burns:2019zia} is illustrated as the green band in the right plot, showing that proposed space-based 21-cm experiments would be sensitive to such injections scenarios. The brightness temperature for the Standard Model expectation (dashed grey) and a phenomenological power law modification to the CMB temperature fit to low-frequency radio surveys (dotted grey) are shown for comparison.}
\label{fig:T21}
\end{figure*}

The 21-cm brightness temperature corresponding to these scenarios is shown in Fig.~\ref{fig:T21}. 
The left panel shows a signal with a spectral endpoint at $z\simeq15$ (with parameters as in Benchmark 1, red line) and lowered X-ray heating alongside the tentative EDGES measurement (blue band). The sharp turn-off in the absorption feature is now predominantly due to the spectral endpoint. For comparison, the case of power law photon injection is shown, with the turn-off due to X-ray heating. Appendix~\ref{app:edges-parameter-space} further explores the viable parameter space within the model considered here that could contribute to the absorption feature observed by EDGES\@. 

The right panel shows the effect of an injection around $z\sim95$ and a kinematic endpoint at $z\sim65$, corresponding to our Benchmark 2 parameter point. The 15\,mK uncertainty projected by the proposed DAPPER experiment in the 15--40\,MHz frequency range~\cite{Burns:2019zia} is shown as the green band around the expected signal. It can be seen that such a distinctive spectral feature would be observable by future 21-cm experiments and easily distinguished from astrophysical backgrounds, providing a new probe of the nature of the dark sector.

\noindent
{\bf Conclusions.---}We have introduced a qualitatively new class of global 21-cm signatures resulting from interactions between the standard and dark sectors, characterized by spectral features---edges and endpoints---and excesses in the observed 21-cm global signal brightness temperature. We have shown how features resulting from dark photon-to-photon conversion can modify the 21-cm absorption trough during cosmic dawn, providing a potential explanation for the anomalous depth and shape of the 21-cm absorption feature measured by EDGES and, more generally, a new way to look for new physics in current and upcoming 21-cm measurements targeting the cosmic dawn era. We have additionally demonstrated how resonant photon injection can result in distinctive spectral features in the brightness temperature during the dark ages, which can be targeted by proposed space-based 21-cm experiments~\cite{Alvarez:2019pss, Burns:2019zia, Koopmans:2019wbn}. 

Although we have focused on a particular model realization here, we emphasize the generality of the signatures introduced. Any exotic resonant photon injection---such as due to conversions between SM photons and axion-like particles~\cite{Moroi:2018vci, Choi:2019jwx}---may generically result in a spectral edge in the 21-cm temperature.   A kinematic endpoint in the model will correspondingly produce a spectral endpoint, which may be hard---as in the case of two-body decay considered here---or soft, as expected for three (or more)-body decay. 

We have also focused exclusively on signatures in the \emph{global} 21-cm signal; the inhomogeneous nature of resonant photon injection~\cite{Caputo:2020rnx} implies that striking signatures may be expected in the 21-cm power spectrum as well, which is expected to be targeted by ongoing and proposed surveys. We defer these additional applications of our framework to future work. 
The code used to obtain the results in this Letter is available at \url{https://github.com/smsharma/edges-endpoints-21cm}\href{https://github.com/smsharma/edges-endpoints-21cm}~\githubmaster.

\vspace{.3cm}

\noindent
{\bf Acknowledgements.---}We thank Yacine~Ali-Ha\"{i}moud, Misha~Ivanov, Julien~Lesgourgues, Julian~Mu\~{n}oz, Josef~Pradler, Roman~Scoccimarro, and Tejaswi~Venumadhav for helpful conversations. AC acknowledges support from the ``Generalitat Valencian'' (Spain) through the ``plan GenT'' program (CIDEGENT/2018/019), as well as national grants FPA2014-57816-P, FPA2017-85985-P, and the European projects H2020-MSCA-ITN-2015//674896-ELUSIVES\@. HL is supported by the DOE under contract DESC0007968. SM and JTR are supported by the NSF CAREER grant PHY-1554858 and NSF grant PHY-1915409. SM is additionally supported by the Simons Foundation.  JTR is also supported by an award from the Alexander von Humboldt Foundation and by the Deutsche Forschungsgemeinschaft (DFG, German Research Foundation) under Germany’s Excellence Strategy – EXC 2121 “Quantum Universe" – 390833306.
JTR acknowledges hospitality from  the Aspen Center for Physics, which is supported by the NSF grant PHY-1607611.
MP is supported in part by U.S. Department of Energy (Grant No. desc0011842). 
AU is supported in part by the MIUR under contract 2017\,FMJFMW 
and by the INFN grant SESAMO\@. This work made use of the NYU IT High Performance Computing resources, services, and staff expertise. The authors are pleased to acknowledge that the work reported on in this Letter was substantially performed using the Princeton Research Computing resources at Princeton University which is a consortium of groups including the Princeton Institute for Computational Science and Engineering and the Princeton University Office of Information Technology's Research Computing department. This research has made use of NASA's Astrophysics Data System. This research made use of the \texttt{ares}~\cite{Mirocha:2014faa}, \texttt{astropy}~\cite{Price-Whelan:2018hus,Robitaille:2013mpa}, CAMB~\cite{Lewis:1999bs,Lewis:2002ah}, CLASS~\cite{Blas:2011rf}, COLOSSUS~\cite{Diemer:2017bwl}, \texttt{HyRec}~\cite{AliHaimoud:2010dx}, \texttt{IPython}~\cite{PER-GRA:2007}, Jupyter~\cite{Kluyver2016JupyterN}, \texttt{DarkHistory}~\cite{Liu:2019bbm}, \texttt{matplotlib}~\cite{Hunter:2007}, \texttt{nbodykit}~\cite{Hand:2017pqn}, \texttt{NumPy}~\cite{numpy:2011}, \texttt{seaborn}~\cite{seaborn}, \texttt{pandas}~\cite{pandas:2010}, \texttt{SciPy}~\cite{2020SciPy-NMeth}, and \texttt{tqdm}~\cite{da2019tqdm} software packages. 

\appendix

\section{Global 21-cm signal calculation}
\label{app:21-cm-code}

In this section we describe the simplified prescription we use for computing the spin temperature evolution and corresponding sky-averaged 21-cm brightness temperature, emphasizing the assumed sources and mechanisms of heating. We stress that, given the large uncertainties associated with the contributing astrophysics, our goal here is simply to model the gross features expected in a 21-cm absorption signal in order to enable a qualitative study of the effect of resonant photon injection. Code implementing the described prescription is available at \url{https://github.com/smsharma/twentyone-global}\href{https://github.com/smsharma/twentyone-global}. We have cross-checked elements of our code against the public global 21-cm code \texttt{ares}~\cite{Mirocha:2014faa}.\footnote{\url{https://github.com/mirochaj/ares}}

\subsection{21-cm temperature evolution}

The evolution of the kinetic temperature $T_\mathrm{k}$ is modeled following Refs.~\cite{Venumadhav:2018uwn,Hirata:2005mz,Mesinger:2010ne},
\begin{align}
\label{eq:T_k_diff}
(1+z) \frac{\dd T_{\mathrm{k}}}{\dd z}=&\, 2 T_{\mathrm{k}}-\frac{1}{\left(1+f_{\mathrm{He}}+x_{\mathrm{e}}\right)} \times \nonumber \\ &\left[\mathcal{E}_{\mathrm{Comp}}+\sum_{\mathrm{r}=\mathrm{c}, \mathrm{i}} \mathcal{E}_{\alpha, \mathrm{r}} \frac{J_{\alpha, \mathrm{r}}}{J_{\alpha, 0}}+\mathcal{E}_{\mathrm{CMB}}\right] T_{\mathrm{k}} \nonumber \\ &-\frac{2 \mu m_\mathrm{p} \Gamma_{X}}{3 \rho_\mathrm{b,0} k_\mathrm{B} H}.
\end{align}
Here, $f_{\mathrm{He}} = 0.08$ is the He/H ratio, $x_\mathrm{e}$ the electron ionization fraction, $\mathcal{E}_{\mathrm{Comp}}$ the Compton heating efficiency computed following Ref.~\cite{Ali-Haimoud:2013hpa}, and $\mathcal{E}_{\alpha, \mathrm{\{c, i\}}}$ the Lyman-$\alpha$ heating efficiencies corresponding to distortions sourced by continuum and injected photons, obtained from Ref.~\cite{Venumadhav:2018uwn}, with fluxes $J_{\mathrm{\alpha, \{c, i\}}}$. $J_{\alpha, 0}$ is the flux scale corresponding to a single photon per hydrogen atom. $\mathcal{E}_{\mathrm{CMB}}$ is the heating efficiency corresponding to energy transfer between CMB photons and IGM mediated by Lyman-$\alpha$ emission, computed in Ref.~\cite{Venumadhav:2018uwn} where it was shown to be especially important in the case of additional photon injection such as considered in this study. The final term represents X-ray heating corresponding to a heating rate $\Gamma_{X}$~\cite{Hirata:2005mz}, with mean molecular weight taken to be $\mu = 0.59$ for a fully ionized gas. Specifics of the assumed stellar populations used to compute $J_{\mathrm{\alpha, \{c, i\}}}$ and $\Gamma_{X}$ will be discussed in the following subsections. Note that we leave the redshift dependence implicit on the right hand side terms of Eq.~\eqref{eq:T_k_diff}.

The spin temperature is given by
\begin{equation}
\label{eq:T_s}
T_{\mathrm{s}}^{-1}=\frac{x_{\mathrm{CMB}} T_{\gamma}^{-1}+\tilde{x}_{\alpha} T_{\mathrm{c}}^{-1}+x_{\mathrm{c}} T_{\mathrm{k}}^{-1}}{x_{\mathrm{CMB}}+\tilde{x}_{\alpha}+x_{\mathrm{c}}},
\end{equation}
where $x_{\mathrm{CMB}}=(1-e^{-\tau_{21}}) / \tau_{21}$ and $x_{\mathrm{c}}$ and  $\tilde{x}_{\alpha}$ represent the relative spin-flip rates from atomic collisions and the Wouthuysen-Field effect, respectively. The former is computed following Ref.~\cite{Pritchard:2008da} and the latter following the procedure and fitting formulae in Ref.~\cite{Hirata:2005mz}. The 21-cm optical depth $\tau_{21}$ is computed following Ref.~\cite{Venumadhav:2018uwn}, and the effective color temperature $T_\mathrm{c}$ is obtained using the fitting formulae in Ref.~\cite{Hirata:2005mz}.

We self-consistently account for the evolution of the ionization fraction due to photoheating during reionization~\cite{Mesinger:2010ne} using the semi-analytic model of X-ray--IGM interaction proposed in Refs.~\cite{Furlanetto:2009uf,Ali-Haimoud:2013hpa}. The corresponding differential equation for $\dd x_\mathrm{e}/\dd z$ is solved together with Eqs.~\eqref{eq:T_k_diff} and~\eqref{eq:T_s} using the \texttt{SciPy}~\cite{2020SciPy-NMeth} implementation of the order 5(4) Runge-Kutta method~\cite{dormand1980family} in order to obtain the evolution of $T_\mathrm{k}$ and $x_\mathrm{e}$, setting initial conditions on these to match the baryon temperature and ionization fraction output by CLASS~\cite{Blas:2011rf} (relying on \texttt{HyRec}~\cite{AliHaimoud:2010dx}) at redshifts well before the onset of reionization or exotic energy injection. 

The spin temperature can finally be obtained using Eq.~\eqref{eq:T_s} and the 21-cm brightness temperature contrast as~\cite{Venumadhav:2018uwn,Hirata:2005mz,Furlanetto:2006fs,Madau:1996cs}
\begin{equation}
\Delta T_{\mathrm{b}}=x_{\mathrm{CMB}} \frac{\tau_{21}}{1+z}\left(T_{\mathrm{s}}-T_{\gamma}\right).
\end{equation}

\subsection{Lyman-$\alpha$ emission}

We follow the prescription in Ref.~\cite{Hirata:2005mz} to construct a toy model for Lyman-$\alpha$ emission. Briefly, the Lyman-$\alpha$ flux is given by
\begin{equation}
J_{\alpha}(z)=\frac{(1+z)^{2}}{4 \pi} \sum_{n=2}^{\infty}  P_{n p} \int_{z}^{z_{\max }} \frac{\epsilon\left(\nu_{n}^{\prime}, z^{\prime}\right)}{H(z^{\prime})}  \mathrm{d} z^{\prime}
\end{equation}
where $\epsilon\left(\nu, z\right)$ is the UV source emissivity, describing the number of photons emitted per comoving volume, proper time, and frequency at redshift $z$ and frequency $\nu$. The factor $ P_{n p}$ accounts for the probability of producing a Lyman-$\alpha$ photon after exciting HI to the $np$ configuration; see Ref.~\cite{Hirata:2005mz} for details. The redshift evolution of the source emissivity is assumed to be proportional to the star-formation rate density, $\epsilon(\nu, z) \propto \dot{\rho}_{*}(z)$~\cite{Barkana:2004vb,Hirata:2005mz,Mirocha:2014faa}
with%
\begin{equation}
\dot{\rho}_{*}(z)=f_{*} {\rho}_{\mathrm b, 0} \frac{\dd f_{\mathrm{coll}}(z)}{\dd t},
\end{equation}
where ${\rho}_{\mathrm b, 0}$ is the mean baryon density today, $f_{*}$ is the star-formation efficiency, which we take to be constant, and
where the rate of collapse $f_{\mathrm{coll}}$ is given by 
\begin{equation}
f_{\mathrm{coll}}(z)=\frac{1}{\rho_\mathrm{m}} \int_{M_{\min }}^{\infty} \dd M \frac{\dd n}{\dd \ln M},
\end{equation}
where $\rho_\mathrm{m}$ is the comoving matter density. The halo mass function $\dd n / \dd \ln M$ is modeled used the parameterization from Ref.~\cite{Despali:2015yla} implemented in the code package COLOSSUS~\cite{Diemer:2017bwl}. $M_\mathrm{min}$ corresponds to the threshold minimum mass of haloes with efficient star formation, parameterized by a cut on the corresponding virial temperature $T_\mathrm{vir}$. The source emissivity is given by
\begin{equation}
\epsilon(\nu, z)=\epsilon_\mathrm{b}(\nu) \frac{\dot{\rho}_{*}(z)}{m_\mathrm{p}},
\end{equation}
where $\epsilon_\mathrm{b}(\nu)$, the number of photons emitted per baryon, is modeled through a blackbody spectrum of temperature $10^5$\,K~\cite{Bromm:2000nn} normalized to an energy-per-baryon of 5.4\,MeV~\cite{Hirata:2005mz}.

\subsection{X-ray heating}

We assume a power-law spectral energy distribution (SED) for the X-ray source population, $I_\nu \propto \nu^\alpha$ with $\alpha = -1.5$, and relate the co-moving X-ray luminosity density $L_X$ to the star-formation rate density $\dot{\rho}_{*}$~\cite{Mirocha:2017xxz,Mirocha:2014faa,Mirocha:2018cih,Schneider:2018xba}:
\begin{equation}
\label{eq:L_X_SFRD}
L_{X}=c_{X} f_{X} \dot{\rho}_{*}(z).
\end{equation}
We assume the canonical value $c_{X} =2.6 \times 10^{39}$ $\mathrm{erg}\,\mathrm{s}^{-1}\left(M_{\odot}\,\mathrm{yr}^{-1}\right)^{-1}$ for the luminosity per star formation rate, normalized in the energy range 0.5--8\,keV~\cite{Mirocha:2017xxz,Mirocha:2018cih} based on extrapolation of the local high-mass X-ray binary (HMXB) luminosity function to higher redshifts~\cite{Mineo:2011id,Mineo:2012qv}, with $f_X$ parameterizing the uncertainty associated with the extrapolation. The assumed SED can then be used to `stretch' the luminosity in Eq.~\eqref{eq:L_X_SFRD} over a wider applicable mass range, for which we take $E_\mathrm{min}=0.2$\,keV~\cite{Mirocha:2017xxz}, in order to obtain the heating rate $\Gamma_X$ used in Eq.~\eqref{eq:T_k_diff}.

Our simplified prescription contains $T_\mathrm{vir}$, $f_*$, and $f_X$ as free parameters, providing the flexibility to regulate an overall heating rate as well as the relative amount of X-ray and UV emission. In the fiducial case, we set $f_X=1$, $f_* = 3\%$, and $T_\mathrm{vir} = 2\times10^4$\,K (corresponding to virial halo mass $M_\mathrm{vir} \simeq 9\times 10^7\,\mathrm{M}_\odot$ at $z=20$).

\section{Parameter-space constraints on EDGES explanation}
\label{app:edges-parameter-space}

\begin{figure}[!htbp]
\centering
\includegraphics[width=0.47\textwidth]{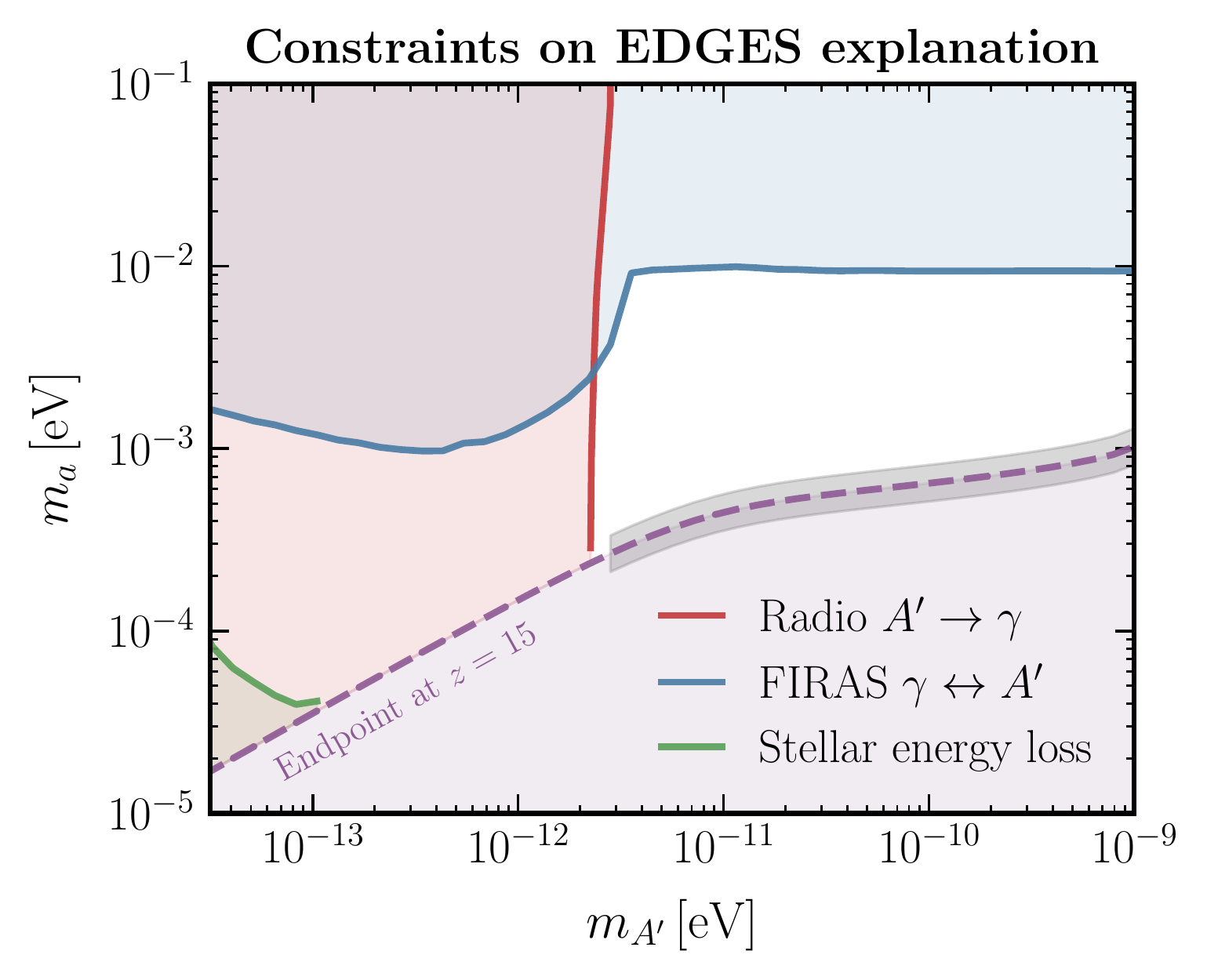}
\caption{Constraints on the parameter space that could explain the anomalous absorption feature observed by EDGES\@. Constraints shown from stellar energy loss due to $A'a$ pair production~\cite{Haft:1993jt,Pospelov:2018kdh} (green region), spectral distortion constraints from $\gamma\leftrightarrow A'$ using COBE/FIRAS~\cite{Mirizzi:2009iz,Caputo:2020bdy,Caputo:2020rnx} (blue region), and constraints on $A'\rightarrow\gamma$ saturating radio observations~\cite{Dowell:2018mdb,Fixsen:2009xn} (red region). In the purple region, produced photons would be too soft to contribute to the EDGES observation. The region in the grey band can simultaneously explain the anomalous depth and sharp endpoint at $z\sim15$ of the measured feature. Mixing parameter values $\epsilon \sim 10^{-6}$--$10^{-8}$ can explain the putative depth of the EDGES observation in the unconstrained part (white region) of parameter space.}
\label{fig:EDGES_param}
\end{figure}

Figure~\ref{fig:EDGES_param} shows the viable parameter space of the model considered in the main body towards explaining the measured EDGES signal~\cite{Bowman:2018yin}.  The model is  constrained by stellar energy loss due to $A'a$ pair production~\cite{Haft:1993jt,Pospelov:2018kdh} (green region), spectral distortions from $\gamma\rightarrow A'$ using COBE/FIRAS~\cite{Mirizzi:2009iz,Caputo:2020bdy,Caputo:2020rnx} (blue region), and the possibility that $A'\rightarrow\gamma$ saturates radio observations~\cite{Dowell:2018mdb,Fixsen:2009xn} (red region). We set the DM lifetime to its saturation limit, $\tau_a = 1.59\times 10^{11}$\,years (95\% confidence level)~\cite{Poulin:2016nat}. Having constrained $\epsilon$ for a given $\mAp$ and $m_a$, we check the maximum allowed 21-cm flux at $z=17$, corresponding to the EDGES observation. The excluded regions correspond to parameter space where the maximum excess 21-cm photon flux is constrained to be smaller than twice the baseline CMB flux, \emph{i.e.}, $(\dd n_\gamma / \dd\omega_{21})|_{z=17} \leq 2 \times (\dd n_\mathrm{CMB} / \dd\omega_{21})|_{z=17}$, which is roughly that required to explain the observed EDGES depth.

Constraints using COBE/FIRAS data were previously considered in Refs.~\cite{Bondarenko:2020moh,Garcia:2020qrp}, which modeled conversions in the presence of plasma mass inhomogeneities at redshifts $z < 6$ using hydrodynamic $N$-body simulations. We present constraints using the complementary semi-analytic approach developed in Refs.~\cite{Caputo:2020rnx,Caputo:2020bdy}, additionally including conversions at higher redshifts where the homogeneous plasma approximation holds. We compute the expected flux in the FIRAS energy bins using
\begin{equation}
\frac{\dd n_{\gamma}}{\dd\omega} = \frac{\dd n_\mathrm{CMB}}{\dd\omega}(1 - \langle P_{\gamma\rightarrow A'}\rangle) + \frac{\dd n_{A'}}{\dd\omega}
\langle P_{A'\rightarrow\gamma}\rangle
\end{equation}
and constrain the mixing parameter $\epsilon$ at the 95\% confidence level using a procedure analogous to the one used in Ref.~\cite{Caputo:2020bdy}. The constraints obtained are shown in Fig.~\ref{fig:EDGES_param} in shaded blue.

Lower-frequency radio surveys further strongly constrain the available parameter space to $m_{A'} \gtrsim 2\times 10^{-12}$\,eV, corresponding to resonant photon injection before $z\gtrsim 330$. These are obtained by requiring that the photon flux due to decay of $a$ to $A'$ and their subsequent oscillation into photons (Eq.~\eqref{eq:dn_domega}) not oversaturate the measured flux above the 2-$\sigma$ level (see the right panel of Fig.~\ref{fig:features} for an example parameter space point that is at the constraint threshold). These constraints, shown in shaded red in Fig.~\ref{fig:EDGES_param}, are generally stronger than those from COBE/FIRAS in the range of dark photon masses $m_{A^\prime}$ probed in Ref.~\cite{Bondarenko:2020moh} using low-redshift ($z < 6$) conversions.

We note that local perturbations in the DM density may be important when the decay of DM $a$ and subsequent conversion of $A'$ are closely separated in time, since this could induce additional spatial correlations in the incoming photon flux. This could be especially important for COBE/FIRAS constraints at masses $\mAp \lesssim 3\times 10^{-12}$\,eV, where decays post-reionization can contribute dark photons which immediately oscillate to photons due to the broad conversion probability at low redshifts~\cite{Caputo:2020rnx,Caputo:2020bdy}. 
We will elaborate more on this point in the next Appendix (see also Ref.~\cite{Bondarenko:2020moh}, which presents COBE/FIRAS constraints while accounting for variations in the local DM density using hydrodynamic simulations), where we show that this effect can be neglected when considering the integrated photon flux from conversions.

The dashed purple line in Fig.~\ref{fig:EDGES_param} corresponds to a kinematic endpoint at $z=15$, which would result in a sharp turn-off of the 21-cm absorption feature. In the purple region, produced photons would be too soft to contribute to the EDGES measurement. The grey band shows the region where resonant dark photon conversion can simultaneously explain the depth and endpoint shape of the putative EDGES signal. In the allowed range of parameter space (white region), mixing coefficient values of $\epsilon \sim 10^{-6}$--$10^{-8}$ are typically required to explain the overall depth of the EDGES measurement within the model considered.

\section{Correlations between decay and resonance conversions}
\label{app:correlations}

For the model considered in the main body of this Letter, both the decay of the DM particle $a$ and subsequent conversion of $A'$ into photons occur in regions of space with densities greater or smaller than the average matter density. As first pointed out in Ref.~\cite{Bondarenko:2020moh}, the densities at the point of decay and conversion may be correlated; if so, this may have a significant impact on the converted photon spectrum, and may affect the limits shown in Fig.~\ref{fig:EDGES_param}. Intuitively, if the points of decay and conversion are separated over cosmological distances, we do not expect any significant correlation to exist. In this appendix, we briefly outline how to make this intuition rigorous, and show that correlations are only important if decay and conversion are separated by a redshift interval $\Delta z$ of less than 1\% of the redshift $z$ at which the decay occurs. 

Ref.~\cite{Caputo:2020bdy} showed that the mean probability of conversion in the presence of inhomogeneities can be calculated by integrating the probability of conversion between $\gamma$ and $A'$ in a region with a density contrast $\delta$ over the probability density function (PDF) of $\delta$.\footnote{We assume for simplicity that DM fluctuations track baryon fluctuations exactly, so that all fluctuations can be described by a single $\delta$, and by the baryon power spectra obtained in Ref.~\cite{Caputo:2020rnx}.} In this model, to get the photon spectrum at some redshift $z$ as a function of $x \equiv \omega/T_\text{CMB}$ in full generality, we must integrate over the joint probability density function of the density contrast $\delta_{\text{dec}}(z_\text{dec})$ at the point of decay and the density contrast $\delta_\text{conv}$ at the point of conversion, with
\begin{alignat}{1}
    1+z_\text{dec} \equiv \frac{m_a(1 + z)}{2 x T_\text{CMB}(z)} \,,
\end{alignat}
where $z_\text{dec}$ is the redshift at which the decay $a \to A'A'$ produces the $A'$ at energy $\omega(z) = x T_\text{CMB}(z)$. Explicitly, the mean photon spectrum produced by $A' \to \gamma$ conversions at redshift $z$ is given by
\begin{multline}
    \frac{\dd}{\dd t} \left( \frac{\dd n_\gamma}{\dd x} \right) (x,z) = \int \dd \delta_\text{dec} \int \dd \delta_\text{conv} \\
    \times \frac{\dd n_{A'}}{\dd x}(x,z) \left[1 + \delta_\text{dec}(z_\text{dec})\right] f\left(\delta_\text{dec}, \delta_\text{conv} \right) \\
    \times \frac{\pi m_{A'}^2 \epsilon^2}{\omega(z)} \left[1 + \delta_\text{conv}(z)\right] \delta_\text{D} \left(1 + \delta_\text{conv} - \frac{m_{A'}^2}{\overline{m_\gamma^2}} \right) \,,
    \label{eq:photon_spec_with_correlations}
\end{multline}
where $f(\delta_\text{dec},\delta_\text{conv})$ is the joint PDF of $\delta_\text{dec}(z_\text{dec})$ and $\delta_\text{conv}(z)$ respectively. The expression for $\dd n_{A'} / \dd x$ is defined in the main body, and depends only on the mean density of DM; the factor of $[1 + \delta_\text{dec}(z_\text{dec})]$ accounts for the actual density of DM at the point of decay. 

Under the assumption that $\delta_\text{dec}$ and $\delta_\text{conv}$ are independent, the PDF $f$ factorizes into the product of their respective one-point PDFs, $f(\delta_\text{dec},\delta_\text{conv}) \to \mathcal{P}(\delta_\text{dec},z_\text{dec}) \mathcal{P}(\delta_\text{conv},z)$. The integral over $\delta_\text{dec}$ is simply equal to 1, while the integral over $\delta_\text{conv}$ reduces to $\dd \langle P_{A' \to \gamma} \rangle / \dd t$~\cite{Caputo:2020bdy}, so that $\dd/\dd t(\dd n_\gamma/\dd x) = \dd n_{A'} / \dd x \times \dd \langle P_{A' \to \gamma} \rangle / \dd t$, which is the expression used in the main body. 

We first specialize to the linear regime, where perturbations are Gaussian, to obtain an expression for $f$. In a matter-dominated universe, perturbations grow linearly with the scale factor, and we can relate $\delta_\text{dec}$ at redshift $z_\text{dec}$, to its value at $z$, where it has grown to a density contrast of $(1+z_\text{dec})/(1+z) \times \delta_\text{dec}$. Furthermore, the point of decay and the point of conversion are separated by comoving distance
\begin{alignat}{1}
    r(z_\text{dec}, z) = \int_{z_\text{conv}}^z \frac{\dd z}{H(z)} \,.
\end{alignat}
This allows us to relate $f(\delta_\text{dec},\delta_\text{conv})$ to the joint probability density function of fluctuations at the same redshift $z$, separated by comoving distance $r$, which in the linear regime is simply a bivariate Gaussian distribution with a two-point correlation function $\xi(z,r)$. Concretely, in the linear regime, the joint PDF is $f_\text{G}(\delta_\text{dec},\delta_\text{conv})$, where
\begin{multline}
    f_\text{G}(\delta_\text{dec}, \delta_\text{conv}) = \frac{1+z_\text{dec}}{1 + z} \frac{1}{2 \pi \sqrt{\left|\mathbf{\Sigma}(z,r)\right|}} \\
    \times \exp \left[-\frac{1}{2} \vec{\delta}^{\,\top} \, \mathbf{\Sigma}^{-1}(z,r) \, \vec{\delta}\right] \,,
    \label{eq:Gaussian_2pt_pdf}
\end{multline}
with the following definitions:
\begin{alignat}{1}
    \vec{\delta} = \begin{pmatrix}
        \frac{1+z_\text{dec}}{1 + z} \delta_\text{dec} \\[6pt] \delta_\text{conv}
    \end{pmatrix}\,, \quad \mathbf{\Sigma}(z,r) = \begin{bmatrix}
        \sigma^2(z) & \xi(z,r) \\[6pt]
        \xi(z,r) & \sigma^2(z)
    \end{bmatrix}\,,
\end{alignat}
together with
\begin{alignat}{1}
    \xi(z,r) = \frac{1}{2\pi^2} \int_0^\infty \dd k \, k^2 P(k,z) \frac{\sin(kr)}{kr} \,,
    \label{eq:two_pt_correlation_fcn}
\end{alignat}
and $\sigma^2(z) = \lim_{r \to 0} \xi(z,r)$. $|\mathbf{\Sigma}|$ denotes the determinant of the covariance matrix, $\mathbf{\Sigma}$. Note that the factor of $(1+z_\text{dec})/(1 + z)$ correctly normalizes $f_\text{G}$ when integrating over $\delta_\text{dec}$ and $\delta_\text{conv}$. Eq.~\eqref{eq:two_pt_correlation_fcn} is the usual definition of the two-point correlation function, with $P(k,z)$ being the power spectrum of fluctuations at redshift $z$, which we can take to be the power spectrum in the linear regime. 

With the bivariate Gaussian distribution in Eq.~\eqref{eq:Gaussian_2pt_pdf}, Eq.~\eqref{eq:photon_spec_with_correlations} can be integrated exactly to give
\begin{multline}    
    \frac{\dd}{\dd t} \left(\frac{\dd n_\gamma}{\dd x}\right) = \frac{\dd n_{A'}}{\dd x} \frac{\dd \langle P_{\gamma \to A'} \rangle}{\dd t} \\ 
    \times \left[1 + \left( \frac{m_{A'}^2}{\overline{m_\gamma^2}(z)} - 1\right) \eta(z_\text{dec}, z) \right] \,,
    \label{eq:photon_spec_linear_regime_integrated}
\end{multline}
where
\begin{alignat}{1}
    \eta(z_\text{dec}, z) \equiv \frac{1+z}{1 + z_\text{dec}} \frac{\xi(z,r)}{\sigma^2(z)} \,.
\end{alignat}

The term $\eta$ compactly expresses the effect of correlations on the converted photon spectrum. In the limit where $\xi(z,r) \to 0$, \emph{i.e.}\ when there are no correlations, we have $\eta \to 0$ and again recover $\dd/\dd t(\dd n_\gamma/\dd x) = \dd n_{A'} / \dd x \times \dd \langle P_{A' \to \gamma} \rangle/ \dd t$ as required. On the other hand, in the $r \to 0$ limit, we get $z \to z_\text{dec}$, and $\lim_{r \to 0} \xi(z,r) = \sigma^2(z)$, so that $\eta \to 1$, with $m_{A'}^2 / \overline{m_\gamma^2} - 1 = \delta_\text{conv} \to \delta_\text{dec}$; the factor $1 + (m_{A'}^2 / \overline{m_\gamma^2} - 1) \eta$ then precisely becomes $1 + \delta_\text{dec}$, the correction due to the decay occurring in an overdensity or underdensity compared to the mean DM density, with conversion happening immediately after the decay.

To establish the importance of correlations, we numerically compute $\eta(z_\text{dec}, z_\text{dec} - \Delta z)$ for all $z$ (taking $\Delta z > 0$), and find $\Delta z$ such that $\eta(z_\text{dec},z_\text{dec} - \Delta z) = 10^{-2}$ ($\eta$ always decreases with increasing $\Delta z$, since $\xi$ gets smaller). This gives $(m_{A'}^2/\overline{m_\gamma^2} - 1) \eta \lesssim 1$, since we always limit overdensities at the point of conversion to be $1 + \delta_\text{conv} < 10^2$ in our treatment of inhomogeneities~\cite{Caputo:2020bdy}. We find that across all $z$,
\begin{alignat}{1}
    \frac{\Delta z}{z_\text{dec}} \lesssim 0.01 \,.
    \label{eq:delta_z_less_than_1_pct}
\end{alignat}
Physically, this means that correlations are only important for dark photons that convert within approximately 1\% of the redshift of decay, and therefore only affect converted photons at redshift $z$ with frequencies that are within 1\% of the endpoint frequency, $m_a/2$. Since the overall spectrum is almost entirely unaffected by the existence of correlations, we find it reasonable to neglect them in the main body of the Letter in the linear regime. 

In the nonlinear regime, a similar argument can be made that shows correlations remain relatively unimportant. Following Ref.~\cite{Caputo:2020rnx}, we make the phenomenologically motivated assumption that density fluctuations follow a log-normal distribution in the nonlinear regime. Explicitly, the one-point PDF for density fluctuations is~\cite{Caputo:2020rnx}
\begin{multline}
    \mathcal{P}_\text{LN}(\delta;z) = \frac{(1+\delta)^{-1}}{\sqrt{2\pi \tau^2(z)}} \\
    \times \exp \left(- \frac{[\ln(1 + \delta) + \tau^2(z)/2]^2}{2 \tau^2(z)}\right) \,,
\end{multline}
where $\tau^2(z) = \ln(1 + \sigma^2(z))$.

To model the growth of fluctuations as a function of redshift from $\delta_\text{dec}$ at $z_\text{dec}$ to $\delta$ at $z$, we use the following prescription: 
\begin{multline}
    \ln(1 + \delta(z)) = \frac{\tau(z)}{\tau(z_\text{dec})} \ln (1 + \delta_\text{dec}(z_\text{dec})) \\
    + \frac{1}{2} \left[\tau(z) \tau(z_\text{dec}) - \tau^2(z)\right] \,.
\end{multline}
This choice enforces $\mathcal{P}_\text{LN}(\delta_\text{dec};z_\text{dec}) \, \dd \delta_\text{dec} = \mathcal{P}_\text{LN}(\delta;z) \, \dd \delta$, so that probability is conserved when mapping between $\delta_\text{dec}$ and $\delta$. Note that in the linear regime, $\ln(1 + \delta) \simeq \delta \ll 1$, and $\tau(z)/\tau(z_\text{dec}) \simeq (1+z_\text{dec})/(1+z)$, with $\tau \sim \delta$ for probable values of $\delta$, so that $\delta$ approximately grows linearly with the scale factor as required. 

The joint PDF for density fluctuations $\delta_\text{dec}$ and $\delta_\text{conv}$, separated by comoving distance $r$, under the log-normal assumption is
\begin{multline}
    f_\text{LN}(\delta_\text{dec}, \delta_\text{conv}) = \frac{\tau(z)}{\tau(z_\text{dec})} \frac{1}{1 + \delta_\text{dec}} \frac{1}{1 + \delta_\text{conv}}  \\
    \times \frac{1}{2 \pi \sqrt{|\mathbf{C}(z,r)|}} \exp \left[- \frac{1}{2} \vec{s}^{\,\top} \, \mathbf{C}^{-1}(z,r) \, \vec{s} \right] \,,
    \label{eq:joint_PDF_LN}
\end{multline}
where
\begin{alignat}{1}
    \vec{s} = \begin{pmatrix}
        \frac{\tau(z)}{\tau(z_\text{dec})} \ln (1 + \delta_\text{dec}(z_\text{dec}))
    + \tau(z) \tau(z_\text{dec}) / 2 \\[6pt] \ln(1 + \delta_\text{conv}) + \tau(z)^2/2
    \end{pmatrix}\,,
\end{alignat}
and
\begin{alignat}{1}
    \mathbf{C}(z,r) = \begin{bmatrix}
        \tau^2(z) & \zeta(z,r) \\[6pt]
        \zeta(z,r) & \tau^2(z)
    \end{bmatrix}\,.
\end{alignat}
Here, the two-point correlation function appearing in $f_\text{LN}$ is
\begin{alignat}{1}
    \zeta(z,r) \equiv \ln \left(1 + \xi(z,r) \right) \,,
\end{alignat}
with $\lim_{r \to 0} \zeta(z,r) = \ln(1 + \sigma^2(z)) = \tau^2(z)$. The fact that $\zeta(z,r)$ correctly gives the two-point correlation function for $s$ in real space, $\langle s(\vec{x}) s(\vec{0}) \rangle$, given the two-point correlation function $\xi(z,r)$ for $\delta$, can be derived from arguments laid out in Refs.~\cite{Greiner:2013jea,Caputo:2020rnx}. Yet again, Eq.~\eqref{eq:photon_spec_with_correlations} can be integrated analytically with $f_\text{LN}$ in Eq.~\eqref{eq:joint_PDF_LN} to give
\begin{alignat}{1}
    \frac{\dd}{\dd t} \left(\frac{\dd n_\gamma}{\dd x}\right) = \frac{\dd n_{A'}}{\dd x} \frac{\dd \langle P_{\gamma \to A'} \rangle}{\dd t} \left[1 + \beta(z_\text{dec},z)\right]\,,
\end{alignat}
where
\begin{alignat}{1}
    \beta(z_\text{dec},z) = \exp \left[ \frac{s_0 \tau(z_\text{dec})}{\tau^3(z)} \zeta(z) - \frac{\tau^2(z_\text{dec})}{2 \tau^4(z)} \zeta^2(z) \right] - 1\,,
\end{alignat}
with $s_0 \equiv \ln \left(m_{A'}^2 / \overline{m_\gamma^2} \right) + \tau^2/2$. We can observe that in both the $\zeta \to 0$ and $r \to 0$ limits, we get $\beta \to 0$ and $\beta \to \delta_\text{dec}$ respectively, as we did in the linear regime. As before, we can assess if correlations are important by finding $\Delta z$ such that $\beta(z_\text{dec},z_\text{dec} - \Delta z) = 1$ for $m_{A'}^2/\overline{m_\gamma^2} = 10^2$. We find once again that Eq.~\eqref{eq:delta_z_less_than_1_pct} is satisfied, therefore allowing us to neglect correlations even in the nonlinear regime. 

\bibliographystyle{apsrev4-1}
\bibliography{features-21cm}

\end{document}